\documentclass{llncs}
\usepackage[utf8]{inputenc}
\usepackage[english]{babel}
\usepackage{booktabs}
\usepackage{makecell}
\usepackage{hyperref}
\usepackage{listings}
\usepackage{xcolor}
\usepackage[nounderscore]{syntax}
\lstdefinelanguage{JavaScript}{
  keywords={break, case, catch, class, const, continue, debugger, default, delete, do, else, export, extends, false, finally, for, function, if, import, in, instanceof, let, new, null, return, super, switch, this, throw, true, try, typeof, var, void, while, with, yield},
  sensitive=true,
  comment=[l]{//},
  morecomment=[s]{/*}{*/},
  morestring=[b]',
  morestring=[b]",
  morestring=[b]`
}
\lstdefinelanguage{WIT}{
  keywords={package, world, interface, import, export, use, include, as, from, with, type, record, variant, enum, flags, resource, func, constructor, static, async, borrow, own, list, option, result, tuple, future, stream, error-context, bool, char, string, u8, u16, u32, u64, s8, s16, s32, s64, f32, f64},
  sensitive=true,
  alsoletter={-},
  comment=[l]{//},
  morecomment=[s]{/*}{*/}
}

\title{AWS Serverless Infrastructure from WebAssembly Components}
\author{Philipp Vollmer}
\institute{\email{briefe@kebes.de}}

\begin{document}
\maketitle

\begin{abstract}
  The infrastructure definitions and the code for serverless
  applications are often only loosely connected.
  This yields a large error class: the referenced infrastructure
  components might not exist. It is also hard to keep
  the application and its permissions in sync.
  To address this problem we present an infrastructure from code
  solution where we provision cloud infrastructure from
  Wasm Interface Type (WIT)---the language defining
  the interface of WebAssembly components.
  We limit ourselves to the basic example of an AWS
  Lambda function accessing S3 buckets.
  We build two interpreters for the WIT language:
  one that derives the definition of S3 buckets as infrastructure as code
  and least-privilege IAM permissions,
  and a generic runtime that runs the component
  as a Lambda function and provides access to the S3 buckets
  as capabilities.
  We formulate consistency properties
  that ensure the existence of the S3 buckets and correct
  permissions of the Lambda, and provide evidence that
  our interpreters satisfy them.
  We show how selected
  failures for deployment of such cloud components
  can be prevented using our technique and
  how a baseline setup fails at this task.
  Together with the strict sandboxing of the
  WebAssembly runtime we achieve
  an instance of capability-based security for access
  to AWS infrastructure as a further result.
\end{abstract}

\section{Introduction}
There are good reasons to adopt serverless architectures
in the cloud: they are very scalable and provide
prepared components for different scenarios: queues,
storage, databases, and functions. They are popular for low-volume
applications as they free developers from the burden of
maintaining server operating systems. Moreover, they
are usually billed ``pay per use'' which makes sense
for applications that are used only occasionally.
They are also popular
for applications with high, variable load because scaling
the system is automatic to some extent.

Yet, developing serverless software in the cloud comes
with engineering challenges.
The more components a piece of software has, the more likely it is
to fail after a change---unnoticed by tests.
Such systems
consist of a lot of interdependent code components which are---from
the code perspective---unrelated.
It is hard to reason about changes in such software systems upfront---for
human developers and coding agents alike. Hence, changes---even
simple ones---must go through several deployments. This costs time,
tokens, and concentration.

We are going to study a very simple example of a serverless
application: an AWS Lambda that accesses a set of S3 buckets.
A setup for deploying such an application
can look as follows:
\begin{itemize}
  \item an npm package containing the code for the Lambda handler
  \item an npm package containing the CDK infrastructure as code
        referencing the first package.
\end{itemize}

The infrastructure as code package will set up the following:
\begin{itemize}
  \item the S3 buckets that the Lambda handler will use
  \item the Lambda function referencing the Lambda handler
        build artifact (usually a CommonJS module)
  \item IAM permissions to access the S3 buckets
\end{itemize}

The following must now be configured in infrastructure as code
in order for the system to work correctly:
\begin{itemize}
  \item The Lambda needs to know how to reference
        the S3 bucket (usually its name will be passed via
        environment variables or will be fixed by convention)
  \item The execution role of the Lambda function must
        contain the proper policies to access
        the S3 buckets. This can be accomplished in
        CDK by method calls on the S3 bucket object.
\end{itemize}

We will call this setup the \emph{baseline setup}.

The following problems can arise when the
infrastructure as code and the handler code do not
match. These problems fit into the taxonomy of Rahman
et al.\ \cite{rahman2020gangofeight}.

\begin{itemize}
  \item \emph{Bucket Name Misconfiguration:} the Lambda code tries to access
        a non-existing S3 bucket due to drift between infrastructure as code
        and Lambda code.
  \item \emph{Permission Misconfiguration:} the Lambda code does
        not have access to the bucket
        or tries to access a bucket it does not have access to.
\end{itemize}
Both problems can be attributed to the \emph{Configuration Data}
category.

Also, the following problem might occur.
\emph{Access Forging:}
the Lambda code might be allowed to perform operations
that are not intentionally derived but can be derived
from the permissions of the execution role.
This problem belongs to the security category.

\section{Prior Art}

\subsection{Related Work}
Pulumi \cite{pulumi} is an Infrastructure from Code tool. It uses
sophisticated closure serialization to serve as a
universal language for describing serverless infrastructure.
Yet, wiring infrastructure components into code and
providing access are not automatic.

Spin \cite{spin} is a framework that facilitates the deployment of
WebAssembly components to Kubernetes and other platforms.
It allows declaring infrastructure components (e.g.\ databases) as
imports in WIT and configures the components via a
deployment manifest.

wasmCloud \cite{wasmcloud-providers} uses capability providers to
give wasm components access to infrastructure.
Cloudflare Workers bindings \cite{cloudflare-bindings} inject
capabilities declared in a configuration file
into a component.

Ampt \cite{ampt} and Encore \cite{encore} are Infrastructure from Code tools.
Winglang \cite{winglang} is a unified programming language for infrastructure
and code.

Table~\ref{tab:related-work} compares these solutions.

\begin{table}[ht]
  \centering
  \caption{Comparison of related infrastructure solutions}
  \label{tab:related-work}
  \begin{tabular}{lcccc}
    \toprule
    Product            & \makecell[b]{Language                   \\Agnostic} & \makecell[b]{Single Source\\of Truth} & \makecell[b]{Permission\\Generation} & \makecell[b]{Capability-Based\\Security} \\
    \midrule
    Cloudflare Workers & yes                   & no  & no  & yes \\
    Winglang           & no                    & yes & yes & no  \\
    Ampt               & no                    & yes & yes & no  \\
    Encore             & no                    & yes & yes & no  \\
    wasmCloud          & yes                   & no  & no  & yes \\
    Spin               & yes                   & no  & no  & yes \\
    Pulumi             & partly                & yes & no  & no  \\
    This work          & yes                   & yes & yes & yes \\
    \bottomrule
  \end{tabular}
\end{table}

Our solution has---though for a much smaller feature
set---all three properties.

\subsection{Foundational Technologies}

Much of our work is supported by sophisticated techniques and
tools. We mention them here to make our contribution to
the topic transparent.

\subsubsection{The Rust programming language wasm32-wasip2 Target}

The Rust programming language supports wasm32-wasip2
as a compilation target \cite{rust-wasm32-wasip2}.
This means that WebAssembly components
which are the foundation of our work can be emitted from
Rust code.

\subsubsection{WIT-Bindgen}

WIT-Bindgen \cite{wit-bindgen} is a tool to generate
traits from WIT definitions. It reduces
the creation of a WebAssembly component implementing
the definition of a WIT file to implementing the traits
generated by WIT-bindgen.

\section{Infrastructure from WebAssembly Components}

In this section we will use the interface description
language \emph{Wasm Interface Type} or \emph{WIT}
as a unified language for describing both the infrastructure
and the runtime for the code. Our paradigm will be
\emph{one language---two consistent interpreters}.

The example use case we describe here
is a serverless function that writes to blob storage
in the cloud. The code for this section is on GitHub \cite{infra-from-wit}.

\subsection{WebAssembly}

WebAssembly \cite{haas2017wasm} is a portable, low-level
bytecode format. The semantics is formally defined.
By itself it has a low-level interface
only allowing the exchange of primitive fixed-size types
(which are usually used to transport pointers).
These wasm modules are
often referred to as \textit{core modules}.

The functions exported by wasm modules usually run compiled
ahead of time in
a memory-isolated sandbox. This is an important security feature
\cite{watt2018mechanising,vanhattum2024crocus}. Other important
features are short startup times and relatively low runtime overhead
compared to native execution (see for example \cite{jangda2019notsofast}).

Interaction with the outer world is possible via imported
functions. We will often denote these functions as
\textit{capabilities}.
While usually conceived as a portable way of running native code in the
browser, its purpose goes far beyond this: WebAssembly targets embedded
and cloud computing as well. One popular example is Cloudflare's
wasm workers. The major cloud providers do not directly support WebAssembly
yet. Hence, we have to invest a little here to fill the gap.

\subsection{The WebAssembly Component Model}
The WebAssembly Component Model \cite{menon2025component,wasm-component-model}
puts one layer of
abstraction on top of wasm core modules. It allows
WebAssembly Components to import and export
functions using a rich type system with heap-allocated
data structures like lists or strings and complex
types like algebraic data types.
It is the basis for all WASI versions from preview 2 on.
WASI \cite{wasi02,wasi-repo}
is a system interface for WebAssembly Components
comparable to POSIX (though at a much higher level).

The interface of WebAssembly modules can be defined using
the WIT language \cite{wit}.

\subsection{Defining the Application using the Component Model}

The component interface definition is given by WIT as follows.
\begin{lstlisting}[language=WIT, caption={The serverless demo application}]
package demo:uploader;

world uploader {
    export aws:lambda/handler@1.0.0;

    @external-id("wasm-capability-demo-uploads")
    import uploads: aws:s3/blob@1.0.0;

    @external-id("wasm-capability-demo-archive")
    import archive: aws:s3/blob@1.0.0;
}

\end{lstlisting}

The interface ``blob'' is defined as follows.

\begin{lstlisting}[language=WIT, caption={The "blob" capability}]
package aws:s3@1.0.0;

interface blob {
    put-file: func(content: list<u8>, name: string);
}

\end{lstlisting}

The interface of a Lambda handler is that of an HTTP
request. We do not repeat it here as it does not contribute
to our discussion.

Most of the language is self-explanatory. A world is
the set of functions a component exports and imports.

\emph{We are going to use the WIT file and the WIT from
  the component interchangeably:} the reason is that
unlike C header files, the interfaces from WIT
can be essentially reconstructed using
the wasm-tools crate \cite{wasm-tools}:

\begin{lstlisting}[language=bash, numbers=none]
wasm-tools component wit <filename>
\end{lstlisting}

\subsection{Deriving the Infrastructure from WIT}

This will be our first interpreter: it reads WIT
and generates CDK from it.
The infrastructure is derived as follows. We start
with the derivation of S3 buckets.
\begin{itemize}
  \item Search for interfaces of the special type aws:s3/blob@1.0.0
  \item Determine the configuration (here given by an external id)
  \item Create the required CDK constructs for the storage.
\end{itemize}
Then we derive the Lambda handlers from it.
\begin{itemize}
  \item Search for interfaces of the special type aws:lambda/handler@1.0.0
  \item Create the required CDK constructs for the handler.
  \item Give the Lambda handler's execution role the required
        permissions on the S3 buckets.
\end{itemize}

Moreover, there will be the need for a runtime
that is not specific to the component. Thus we need to
\begin{itemize}
  \item Build a zip file with the runtime and the component.
  \item Declare an AWS Lambda using this zip file.
\end{itemize}

\subsection{The Runtime}
This is our second interpreter.
In the code it is the \texttt{runtime} crate.
It searches for an entry point
that can be used as the Lambda handler.
Wasm Components have no capabilities by default: they
run as functions inside a sandbox without I/O
capabilities.
Our runtime acts as a dynamic linker. It searches
for imported capabilities in the module. Then
it assembles a function ``put-file'' that
puts a file to the S3 bucket given by the external id
and links it to the WebAssembly component.

\section{Evaluation}

We will now compare our approach with the baseline
setup.
There are three stages at which failures can be detected:
\begin{enumerate}
  \item Prevented by construction: an error
        is made impossible by the way
        our infrastructure is constructed.
  \item Upon local execution: the issue gets detected
        when locally executing tests.
  \item At runtime, deployed in the cloud.
\end{enumerate}

The second one deserves explanation: the runtime can
be imported as a function and can run components.
The S3 capabilities will then be provided as
in-memory implementations. This test harness
is a good simulation of a cloud deployment
and we will mean this test harness when we refer
to local tests.

\subsection{Correctness Properties}

We can now make our correctness
property
\emph{one language---two consistent interpreters} precise.
By this we mean the following:
\begin{itemize}
  \item For every import $i$ with external id $e$ of a
        component $c$,
        an S3 bucket with name $e$ is created
        and a function $f$ that only accesses $e$ is
        provided to $c$. We denote these
        functions by $\texttt{cap}(c)$.
  \item For every function $f$ in
        $\texttt{cap}(c)$ and every AWS action
        $a$ that $f$ calls, we generate an IAM
        statement allowing this action and these
        are the only IAM actions generated.
\end{itemize}

The first property is backed by tests. The second
property comes from the fact that the
implementation of the capability and the code
generating the IAM actions live in the same module.
There is a test in
\path{extractor/src/kinds/s3_blob.rs}
that records all S3 actions in the capability passed
to the component and checks whether the required
IAM grants are generated. The test passes;
given that the recorded actions
map onto the IAM actions required to
grant permission, this provides evidence for the
second property.

However, we clearly
state that this consistency property is a
requirement for the validity of our approach.

\subsection{Access Forging}
The forging of access to infrastructure components
usually works along the following lines:

\begin{itemize}
  \item get credentials from instance metadata service
  \item get credentials from environment variables
  \item create an authenticated AWS SDK client
  \item call an API
\end{itemize}

Interestingly, the STS get-caller-identity endpoint
does not need permissions and cannot be forbidden
in the execution role.
Lambda functions can use the described way to call
APIs that they were not supposed to call.

We have run explicit test cases for the runtime
to verify that each of these steps fails
for our runtime. The component has neither TCP nor
environment variable access.

Our access control to S3
buckets is an implementation of \emph{capability-based security}:
The access to resources cannot be forged from ambient
authority and has to be passed explicitly as a handle
(in this case: a function) to the component.

As a corollary, we see that even forceful attempts
to access S3 buckets that were not explicitly
configured as capabilities are unsuccessful.

We have tested the impossibility of accessing AWS
APIs in the \texttt{runtime/test} folder for our solution.
We have simulated the access forging for the baseline
in the \texttt{thinAir} example in the \texttt{pitfalls}
folder.

\subsection{Bucket Name Misconfiguration}
Due to the lack of network connectivity, the
only way for a component to access S3 buckets
\emph{at all} is by importing the interfaces supplying
access to the S3 buckets. The bucket that is accessed
using the interface is completely abstract; the configuration
is invisible. So this failure type is impossible
\textit{by construction}.
Even attempts that do not use the provided capability
but try to find credentials and use them to access the
AWS API are prevented due to lack of network connectivity
in the sandbox attempts will already fail
on \emph{local execution} of the comonent.
For the baseline, the \texttt{EnvMismatch} example in the \texttt{pitfalls}
directory shows how this case passes compilation
and synthesis and only fails at run time. There
are tests in \path{pitfalls/lambda/test/handlers.test.ts}
for the S3 access operations but the
responses from the S3 service are simulated by a test double.
These tests pass but the access fails when deployed in the cloud.

\subsection{Permission Misconfiguration}
The required infrastructure permissions are set up
automatically by the extractor.
So this failure type is impossible
\textit{by construction}.
The \texttt{MissingGrant} example in the \texttt{pitfalls}
directory shows how this case passes compilation
and synthesis and only fails at run time in the cloud.
There are also tests \path{pitfalls/lambda/test/handlers.test.ts}
which are successful but the example fails in the cloud.

\subsection{Summary of Failure Types}
We see that we are able to detect failures before the
first deployment that, in the baseline setup, will
only surface after a deployment.

\begin{table}[ht]
  \centering
  \caption{Failure types and the stage at which they are detected}
  \label{tab:failure-detection}
  \begin{tabular}{lcc}
    \toprule
    Failure Type                 & Our solution & Baseline            \\
    \midrule
    Bucket Name Misconfiguration & 1 or 2       & 3                   \\
    Permission Misconfiguration  & 1 or 2       & 3                   \\
    Access Forging               & 2            & succeeds undetected \\
    \bottomrule
  \end{tabular}

  \smallskip
  {\footnotesize 1 -- by construction \quad 2 -- in local tests \quad 3 -- deployed in the cloud\par}
\end{table}

\subsection{Performance}
We also want to point out some performance implications.
Depending on the requirements, the impact might be
significant. There is one additional optimization we included
in the benchmarks. Wasmtime compiles wasm modules
ahead-of-time. It also allows saving and loading
a compiled artifact. We add this
as an additional data point.
Our test setup is shown in Table~\ref{tab:benchmark-environment}.
All benchmarks can be found in the code for this article.

\begin{table}[ht]
  \centering
  \caption{Benchmark environment}
  \label{tab:benchmark-environment}
  \begin{tabular}{lp{0.7\linewidth}}
    \toprule
    CPU       & Intel Core i9-8950HK (6 cores, 12 threads), 32\,GB RAM             \\
    OS        & Linux 6.17.0                                                       \\
    Rust      & 1.99.1; native binaries for \texttt{x86\_64-unknown-linux-musl},
    components for \texttt{wasm32-wasip2}                                          \\
    Libraries & wasmtime 50.0.0, lambda\_http 1.3.2, tokio 1.54.1, criterion 0.7.0 \\
    Date      & 2026-11-05                                                         \\
    \bottomrule
  \end{tabular}
\end{table}

\subsubsection{Cold Start Times}

These are the results of a small benchmark measuring
the cold start penalty we get when we work with
WebAssembly and pre-compiled WebAssembly.
In Table~\ref{tab:cold-start} we find the result of
a benchmark with 50 samples. Note that
we have not included the contributions from the AWS Lambda
service in our benchmark.
We can draw three conclusions here.
\begin{itemize}
  \item cold start times are significantly longer when using
        WebAssembly compared to native binaries
  \item cold start times can be improved by a factor of 8
        when pre-compiling the wasm module
  \item However, compared to the cold start times
        \emph{including} the contribution from the
        AWS cloud infrastructure (which is about
        20\,ms, cf.\ \cite{lambda-perf}) the advantage is much smaller
\end{itemize}

\begin{table}[ht]
  \centering
  \caption{Cold start times in ms (single-threaded)}
  \label{tab:cold-start}
  \begin{tabular}{lrrrrrr}
    \toprule
                & \multicolumn{2}{c}{p50} & \multicolumn{4}{c}{total}                                 \\
    \cmidrule(lr){2-3} \cmidrule(lr){4-7}
    Scenario    & init                    & invoke                    & min   & p50   & p90   & mean  \\
    \midrule
    native      & 5.0                     & 0.32                      & 2.0   & 5.3   & 7.0   & 5.1   \\
    wasm        & 105.0                   & 0.55                      & 102.7 & 105.6 & 108.0 & 105.8 \\
    precompiled & 12.0                    & 0.52                      & 9.4   & 12.5  & 14.1  & 12.5  \\
    \bottomrule
  \end{tabular}
\end{table}

\subsubsection{Artifact Sizes}

We measure the artifact sizes. We have three different components
here.
\begin{itemize}
  \item \texttt{bootstrap}: the entry point executable
  \item the wasm component
  \item the pre-compiled wasm component
\end{itemize}
We see that the WebAssembly runtime contributes significantly
to the artifact sizes.

\begin{table}[ht]
  \centering
  \caption{Deployment artifact sizes in MB}
  \label{tab:artifact-sizes}
  \begin{tabular}{llrr}
    \toprule
    Scenario    & Deployment zip contents                   & Raw   & Zipped \\
    \midrule
    native      & \texttt{bootstrap}                        & 4.00  & 1.64   \\
    wasm        & \texttt{bootstrap} + \texttt{bench.wasm}  & 45.27 & 15.23  \\
    precompiled & \texttt{bootstrap} + \texttt{bench.cwasm} & 45.41 & 15.27  \\
    \bottomrule
  \end{tabular}
\end{table}

\subsubsection{Performance}

Although WebAssembly gets compiled to native code, it is significantly
slower than native code. At least two reasons explain this:
\begin{itemize}
  \item \emph{Sandboxing:} the strict sandboxing of WebAssembly
        runtimes slows down execution.
  \item \emph{Missing optimizations:} native compilers can make
        optimizations that WebAssembly does not offer. WebAssembly
        has the simd128 proposal, but we did not test it.
\end{itemize}

For our benchmark we ran the \texttt{sort\_unstable} function from
the Rust language standard library on a vector of
size $n$ of 64-bit unsigned integers. We present our results
in Table~\ref{tab:sort-performance}. As a rough estimate
we can state that native execution is 1.5 times faster
than WebAssembly.

\begin{table}[ht]
  \centering
  \caption{Runtime of \texttt{sort\_unstable} on $n$ 64-bit unsigned integers}
  \label{tab:sort-performance}
  \begin{tabular}{rrrr}
    \toprule
    $n$       & native      & wasm        & wasm / native \\
    \midrule
    10,000    & 124\,$\mu$s & 273\,$\mu$s & 2.2$\times$   \\
    100,000   & 1.53\,ms    & 2.56\,ms    & 1.7$\times$   \\
    1,000,000 & 18.1\,ms    & 27.9\,ms    & 1.5$\times$   \\
    \bottomrule
  \end{tabular}
\end{table}

\section{Limitations}
\subsection{Dependency on an Unstable Features of WebAssembly}
Currently, we use the unstable \texttt{external-id} feature
of WIT. It is not intended for declaring infrastructure
dependency configuration. We also depend on the
unstable \texttt{implements} feature but its use
is idiomatic.

\subsection{No Empirical Data about the Likelihood of the Investigated Errors}
Although it appears plausible from the author's personal
experience---errors of this kind occur frequently
in his everyday work---we did not conduct empirical studies
on these errors. One might deny the importance of
such errors altogether or might object that these
are prevented by common best practice.

\subsection{The Runtime is the Security Boundary}
Although it is backed by strict IAM policies for
the execution role, the runtime for the wasm components
is the security boundary. This implies that extra care
has to be taken when developing it.

\subsection{No Configurability of the Bucket Name}
For multi-stage deployments (staging, production) this is a
challenge as bucket names must be globally unique. In
our case this means that for every stage we would
have to build a separate component.
However, there are resolutions for this issue whose
implementation seems feasible.
\begin{itemize}
  \item Use the id only as an identifier that allows the
        runtime and extractor to do the assignment (the bucket
        name gets generated by CDK and injected via an
        environment variable). This would be
        the idiomatic CDK way.
  \item Allow simple templating inside the external id (allow \texttt{stage} as
        a variable inside templates). This would be the idiomatic
        Azure Bicep way.
\end{itemize}

\section{Conclusion}
Although
the idea of facilitating infrastructure by importing
functions is basically everywhere
in the specification and the examples for
WebAssembly Components, there is no clear
integration path (the stringly-typed external-id is
a workaround). The big advantage and the author's main
motivation is that WIT can
act as a universal language for the infrastructure
while the implementation of the
infrastructure components can be written in almost any
other language. Yet the types in WIT are usually checked
to agree with the types of the implementation---at least
if the language for the implementation is
statically typed.
It would
be interesting to enrich WIT with language elements
that have the semantics of configuration of
dependencies. Also our scope was limited
to one serverless function - adding more
types of infrastructure components and
more ways of interaction between them is an interesting
challenge for WIT language design.

\section*{Declaration of AI-Assisted Technologies}

The author hereby discloses the use of generative ai during
the work on this article. The author used Claude 5.5 Opus by Anthropic
for generating the code for this article. The conceptual ideas, choice
of tools, architecture, design, and structure are the author's own.
The tests and benchmarks were performed and supervised by
the author. The author claims full responsibility for the final results.
The article is written by the author with the aforementioned tool
used for review, spell checking, and formatting.

\bibliographystyle{splncs04}
\bibliography{infrafromwit}

\end{document}